# Automatizing a non-scripting TPS for optimizing clinical workflow and re-optimization of IMRT/VMAT plans

## Rafael Ayala, Gema Ruiz and Teresa Valdivielso

Servicio de Dosimetría y Radioprotección, Hospital General Universitario Gregorio Marañón, Madrid, Spain

E-mail: rafael.ayala@salud.madrid.org

## Abstract:

A toolkit for interacting with the Elekta Monaco (Clements *et al* 2018) treatment planning system (TPS) has been designed without the need of a dedicated Application Programming Interface (API). It provides automatization of the radiotherapy planning procedure allowing the TPS to calculate or optimize plans during non-working hours. The software is based on an open source library that mimics human interaction with Microsoft Windows applications.

The impact on the clinical workflow is important not only providing better efficiency but also increasing treatment quality. Successful inverse planning depends on the tweaking of many parameters that can be explored more exhaustively with this tool without significantly increasing planning time.

Furthermore, a simple way to analyze calculated plans and the impact of the cost functions in the optimization has been implemented.

The Autoflow sequence has been developed allowing fully automated planning; it works analyzing the relative impact of different cost functions in the optimization and modifying constraints accordingly. It also changes segmentation parameters to fit more complex treatments.

A prostate study has been conducted, comparing automatically created plans with already treated ones. The Autoflow sequence has proved to reduce dose to organs at risk with a negligible decrease in target coverage.

Implementing this tool has enabled an efficient use of the TPS, allowing research and clinical use to coexist in a friendly way. It is surely of great use in clinics with little resources. It provides consistency and efficiency throughout the treatment planning process.

## 1. Introduction

Intensity-modulated radiation therapy (IMRT) / volumetric-modulated arc therapy (VMAT) is becoming the standard delivery technique for most of the treatments in a radiotherapy clinic. It requires, though, extensive resources, from hardware to treatment planner time. An efficient use of the TPS is needed to ensure proper planning timings. This is more important in clinics where resources are scarce; there is a potential impact in quality of treatments if the search of the optimal plan is reduced due to time constraints. Introducing some automated tasks can have an enormous impact in the clinical workflow but, at the same time, giving some room to research can improve treatment quality in some centers. Inter-planner variation in plan quality is also another important concern, an automatic workflow can help to provide a base level of performance (Nelms *et al* 2012). Efforts have been made in this field with very different approaches: fully automated solutions with gantry optimizations (Breedveld *et al* 2012, Heijmen *et al* 2018), data-driven (Wu *et al* 2011) or recording user interactions with the TPS in an autopilot fashion (Wang and Xing 2016). Our work follows the latter approach; we interact with a commercial TPS externally.

This study has several purposes: (1) Developing a Monaco scripting toolkit: Pymonaco was created to interact with the TPS with a set of functions written in Python language for the sake of simplicity. (2) Creating a centralized calculation queue: a web server was used to build a calculation queue that runs overnight. (3) Programmatically interpret Monaco results: analyze the results of the optimizations (PTV coverage, impact of cost functions in the optimization). (4) Building a complete sequence for automatic planning: modify constraints according to the result of a previous optimization and developing the Autoflow sequence for fully automated treatment planning. (5) Validating the Autoflow sequence with a prostate retrospective study performed comparing automatic plans with already delivered treatments to ensure clinical suitability.

The automatically created plans proved to be at least comparable to the manual ones, a statistical study shows significant reduction in mean dose to organs at risk whilst losing a very small amount of target coverage.

## 2. Materials and Methods

*2.1. Treatment Planning System:*
Monaco version 5.10, Elekta AB (Stockholm, Sweden), has been used in this study, a Microsoft Windows-based treatment planning system. It provides biological IMRT/VMAT inverse optimization by means of the EUD (Niemierko 1997) and g-EUD (1999) concepts. Optimization can be performed in two different modes, pareto and constrained. All the calculations in this study made use of constrained optimization. Calculations were performed with the Monte Carlo algorithm (Fippel *et al* 2003).

This software version does not provide a scripting API, however, development software tools seem to be available within a research agreement with Elekta (Winkel *et al* 2016).

*2.2. Pymonaco*
We developed a Python module named Pymonaco that automates interactions with Monaco. It is based on the open-source library pywinauto (Mahon *et al* 2018). It acts inspecting UI elements from a Windows application and enabling interactions such as keystrokes or mouse clicks.

Pymonaco comprises several Python functions that manage useful user interactions (e.g., opening or closing a patient and calculating or optimizing a treatment plan). There are multiple ways to perform some of the interactions; we chose one over the other with robustness in mind.

As a first approach, we created a web based TPS scheduler that allows us to calculate or optimize plans during non-working hours. A plan has to be created manually from a template and saved with the desired name. The job has to be included in the calculation list via web specifying the station name where it is going to be calculated. A Windows automated task is run on every Monaco station at a specific time, it receives the calculation queue from the web server and starts all the jobs sequentially (Figure 1).

The job list includes an Autoflow setting which is explained later on.

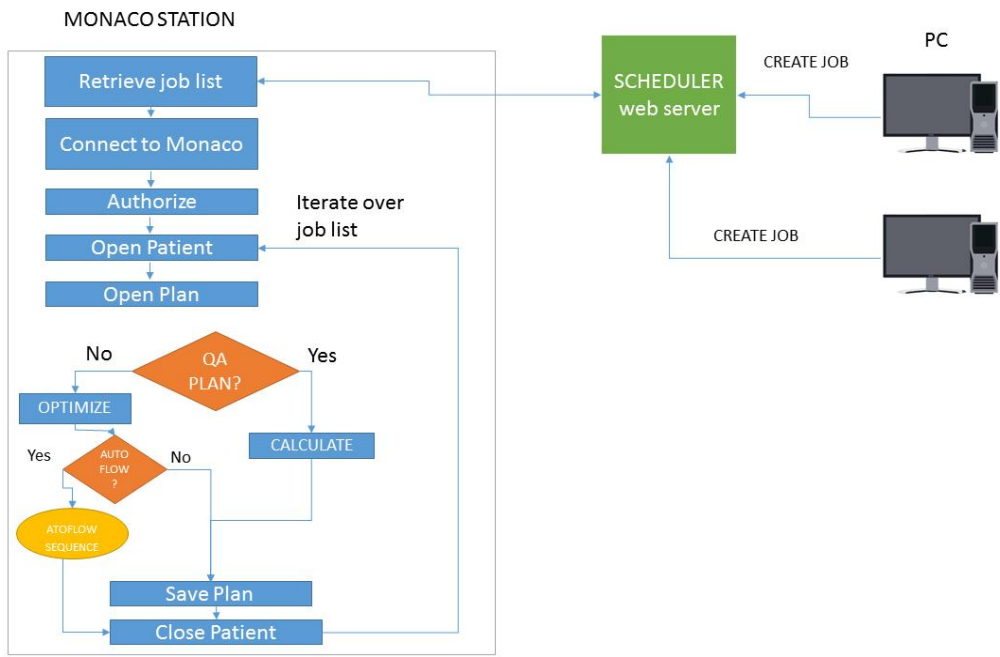

**Figure 1. Treatment planning scheduler description and flowchart.**

## 2.3. Interpretation of calculated plans in Monaco

In Monaco TPS, every plan needs a wish list which contains several cost functions, each of them will have an associated isoconstraint value (Clements *et al* 2018). During the optimization, evaluating the cost functions will generate new values called isoeffects that will be compared to isoconstraints, a value called relative impact will also be calculated to display the importance of the cost function in the optimization. Once the plan is calculated and saved, these final results are written to a text file with extension .hyp. It is then relatively easy to analyze calculated plans and also to adjust cost functions with an external tool. It is also possible to read or modify segmentation parameters such as the degree of fluence smoothing or the maximum number of arcs per beam.

## 2.4. Autoflow

### 2.4.1. General Description

We would like to be able to analyze whether a plan has a good PTV coverage without the need to look at the entire DVH. The isoeffect of the target penalty cost function is already giving us this information. We could evaluate the impact of the cost functions of the OARs and modify the isoconstraints accordingly.

For this purpose, we defined a coverage index (*cv*) as follows:

$$cv = \min\left(\frac{isoeffect(TP_i)}{isoconstraint(TP_i)}\right)$$

Where TPi corresponds to the target penalty cost function of the ieth PTV. We chose a threshold of 0.95 for the cv index that correlates with a good PTV coverage.

Based on the cv index, we built a decision tree to further optimize radiotherapy plans. It is a tradeoff between goodness of plans and calculation time (Figure 2).

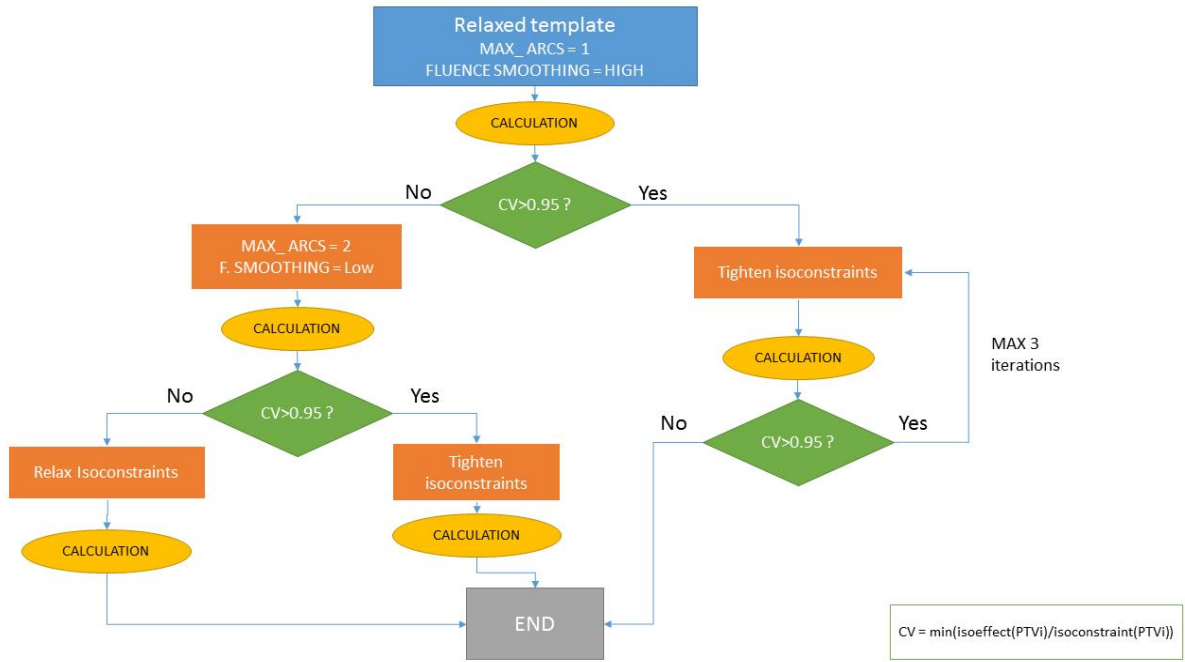

Figure 2. Autoflow sequence flowchart. A maximum of 4 plans are created.

We start with a relaxed template where all the constraints for organs at risk are set to the minimum required to comply with the corresponding protocol. Segmentation settings are also set so that they guarantee a minimum degree of complexity, i.e., one arc per beam and high fluence smoothing. If the resulting calculation has a cv greater than 0.95 it means that there is room for tightening the OAR constraints. This part of the sequence will be performed a maximum of three times if the cv is still over the threshold. If it is not the case, the sequence will stop.

On the other hand, if the initial cv is below 0.95 the plan will be calculated with a higher degree of complexity, increasing the maximum number of arcs per beam to 2 (if the technique is different than VMAT, this parameter will take no effect) and decreasing fluence smoothing to Low. At this point if the cv is higher than the threshold, the isoconstraints will be tightened but just once. On the contrary, if the cv is lower than 0.95 the isoconstraints will be relaxed.

A maximum of 4 plans and a minimum of 2 are therefore calculated during each run.

### 2.4.2. Tightening cost functions
At this step the following OAR cost functions will be modified: Quadratic Overdose, Serial, Parallel and Conformality. Cost functions that correspond to targets will be left untouched as well as Maximum Dose since it works as a hard constraint that will easily degrade the result of the optimization.

The isoconstraints will be changed only if the isoeffect is lower than the isoconstraint increased by 5%. In that case the new isoconstraint will be modified according to the relative impact parameter (Table 1).

| Relative impact | New isoconstraint |
|---|---|
| < 0.25 | Isoeffect * 0.92 |
| [0.25, 0.75) | Isoeffect * 0.95 |
| [0.75, 1) | Isoeffect * 0.98 |
| = 1 | Isoeffect (unchanged) |

Table 1: New isoconstraints based on isoeffect and relative impact.

*2.4.3. Loosening cost functions*

A similar approach than previous is followed for loosening cost functions. For those cost functions where the isoeffect is higher than the isoconstraint, the new isoconstraint will be modified to match the value of the isoeffect increased by 5%.

*2.5. Case studies*

71 VMAT prostate treatment plans have been reoptimized with the Autoflow sequence. They were all treated with an integrated boost technique (SIB) corresponding to a fractionation scheme of 70 Gy delivered to the prostate and 56 Gy to the seminal vesicles along 28 fractions. The simplicity of the plans will allow us to evaluate the validity of the sequence for clinical use. All the plans were calculated with the same machine model, an Elekta linac with Agility MLC and energy of 6MV with flattening filter. The same template was used for all the cases, it comprises one full VMAT beam with a segmentation setting of a maximum of 2 arcs per beam and a fluence smoothing set to the highest possible value (High). Ideally, the input template for the Autoflow sequence should have a maximum of 1 arc per beam, but we realized that most treatments were planned with 2 arcs per beam. The template has very relaxed isoconstraints that will be tightened during the iterative process. Several DVH points were evaluated for PTV, bladder and rectum; conformity (van't Riet *et al* 1997, Feuvret *et al* 2006) and heterogeneity (Quan *et al* 2012) indexes (CI, HI) of the PTV are also reported according to the following definitions:

$$CI = \frac{(Volume\ of\ target\ receiving\ the\ prescription\ dose)^2}{Target\ volume\ *\ Volume\ of\ the\ prescription\ isodose}$$

$$HI = \frac{Dose\ that\ covers\ 1\ \%\ of\ PTV}{Dose\ that\ covers\ 95\ \%\ of\ PTV}$$

*2.6. Statistics*

We have compared the Autoflow sequence and the clinical plans with a two-sided Wilcoxon signed-rank test to assess statistical significance (p<0.05). Scipy (Oliphant 2007) and Matplotlib (Droettboom *et al* 2017) Python modules have been chosen to perform all the analyses and graphical representations.

# 3. Results

*3.1. Monaco scheduler:*

At the time of writing, every Monaco station retrieves its job list and starts optimizing plans at 23:00 every day. All the plans compared in this study have been calculated during non-working hours, thus, not perturbing the clinical workflow.

*3.2. Autoflow:*

The complete Autoflow sequence with 4 plans calculated takes an average of 2.5 hours in an Intel Xeon E5-2695 with 24 cores and 32GB of RAM but can be increased considerably for large target volumes.

A DVH comparison of four successive iterations during the Autoflow sequence of a prostate treatment is shown in Figure 3.

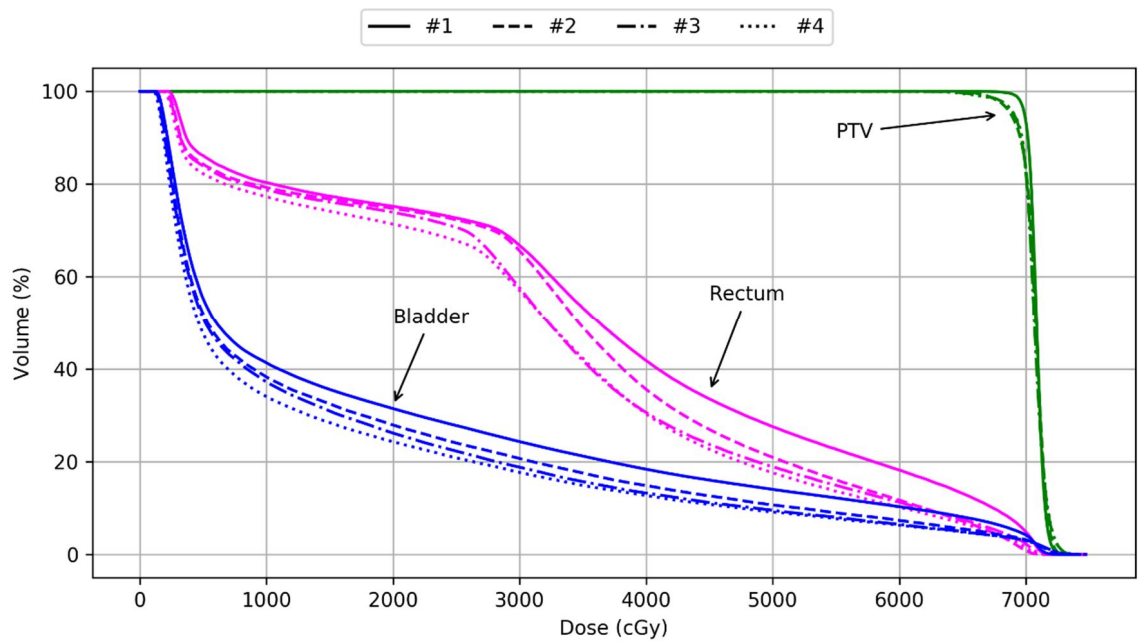

**Figure 3. DVH comparison of a re-optimized plan with the Autoflow sequence. Isoconstraints tightened three times.**

*3.3 Prostate Cases:*

The analysis of the 71 automatically created plans shows total compliance with the current clinical protocols, each treatment plan is, therefore, clinically acceptable.

Autoflow plans were compared to the original plans, which were clinically accepted and already treated. Tables 2 and 3 show histogram differences between them, together with their p-value.

|  | PTV 70Gy | | | | PTV 56Gy | | | |
|---|---|---|---|---|---|---|---|---|
|  | ΔV95% (%) | ΔV107% (%) | ΔHI | ΔCI | ΔV95% (%) | ΔV107% (%) | ΔHI | ΔCI |
| median | -0.17 | 0.00 | 0.00 | -0.04 | -0.03 | 0.73 | 0.00 | -0.03 |
| p-value | 0.018 | 0.077 | 0.051 | 0.004 | 0.342 | 0.023 | 0.026 | <0.001 |

Table 2: PTVs, medians of differences between Autoflow and already treated plans together with their corresponding p-values.

|  | RECTUM | | | | | BLADDER | | | |
|---|---|---|---|---|---|---|---|---|---|
|  | ΔV30Gy (%) | ΔV40Gy (%) | ΔV60Gy (%) | ΔV70Gy (%) | ΔDmean (cGy) | ΔV35Gy (%) | ΔV50Gy (%) | ΔV70Gy (%) | ΔDmean (cGy) |
| median | -3.17 | -0.15 | -1.03 | -0.45 | -128.30 | -7.14 | -2.67 | -0.17 | -451.30 |
| p-value | 0.014 | 0.358 | 0.007 | <0.001 | 0.001 | <0.001 | <0.001 | 0.096 | <0.001 |

Table 3: OARs, medians of differences between Autoflow and already treated plan together with their corresponding p-values.

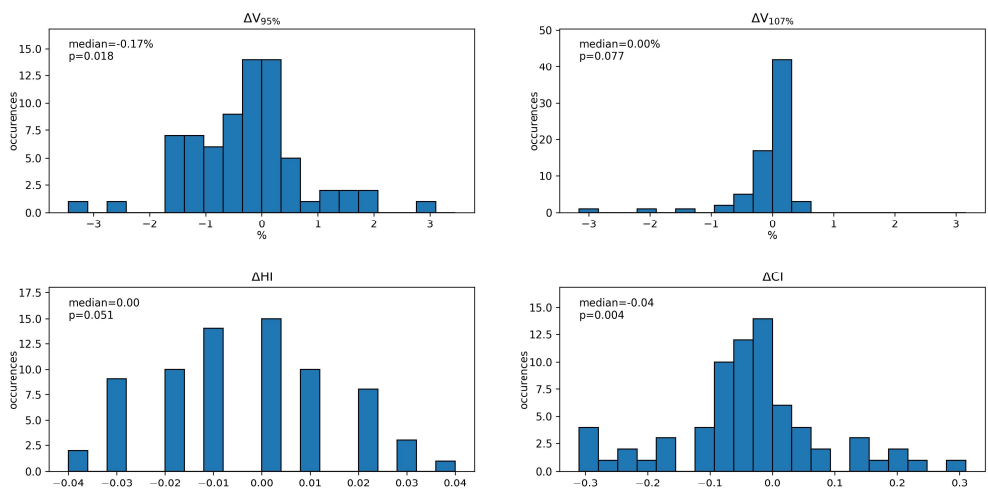

Figure 4. PTV 70Gy ΔV$_{95\%}$ (%), ΔV$_{107\%}$ (%), ΔHI and ΔCI.

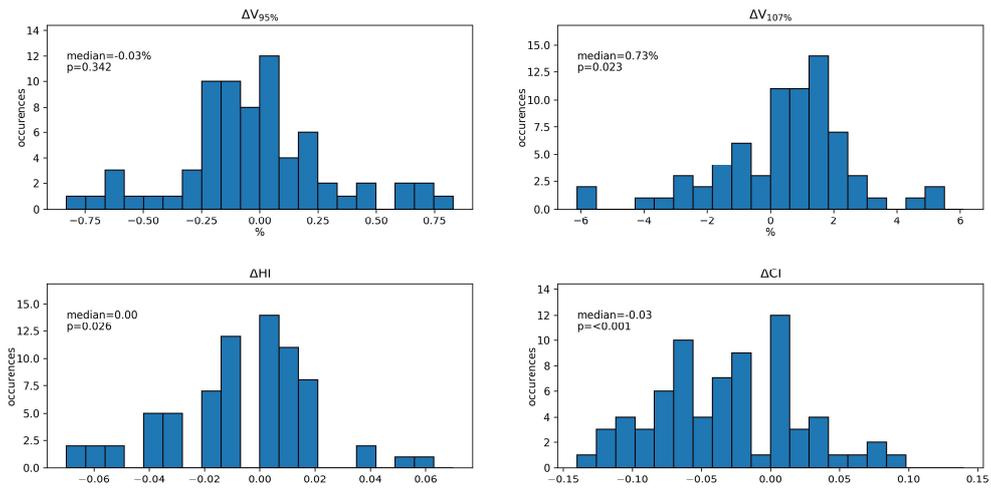

**Figure 5. PTV 56Gy ΔV$_{95\%}$ (%), ΔV$_{107\%}$ (%), ΔHI and ΔCI.**

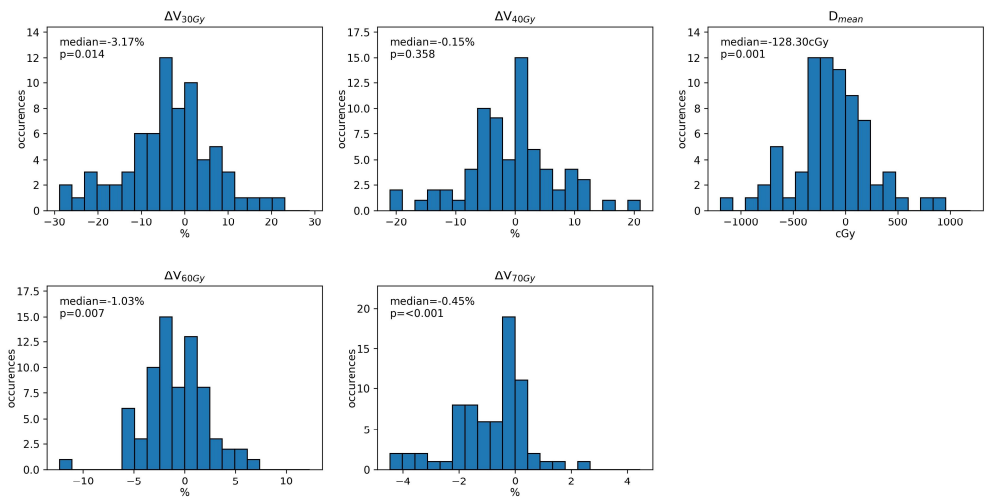

**Figure 6. Rectum ΔV$_{30Gy}$ (%), ΔV$_{40Gy}$ (%), ΔV$_{60Gy}$ (%), ΔV$_{70Gy}$ (%) and D$_{mean}$ (cGy).**

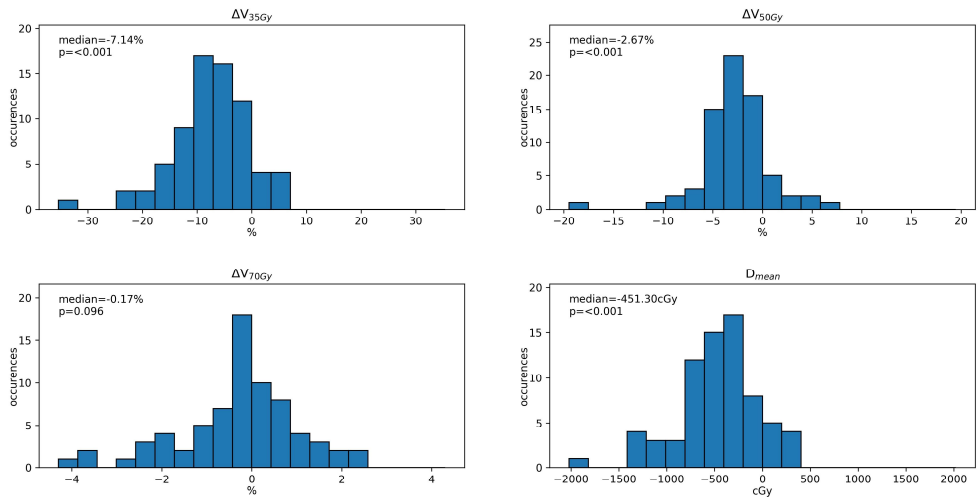

Figure 7. Bladder $\Delta V_{35Gy}$ (%), $\Delta V_{50Gy}$ (%), $\Delta V_{70Gy}$ (%) and $D_{mean}$ (cGy).

The target coverage of PTV 70Gy ($V_{95\%}$) is reduced in less than 0.2% (p=0.018) and hot spots ($V_{107\%}$) are not significantly increased. There is a statistically significant but small decrease in CI (-0.04, p=0.004) and no significant change in heterogeneity index can be seen. (Figure 4)

The analysis of PTV 56Gy shows no significant change in target coverage but a small increase in hot spots ($\Delta V_{107\%}$: 0.73%, p=0.023), no change in heterogeneity index can be observed and CI shows the same behavior as in PTV70Gy (-0.03, p<0.001). (Figure 5)

Regarding OARs, automatic plans significantly reduce low doses to the bladder ($\Delta V_{35Gy}$: -7.14%, p<0.001; $\Delta V_{50Gy}$: -2.67%, p<0.001) whilst not significantly affecting the high dose region. The mean dose to the bladder is drastically reduced (-451.30cGy, p<0.001). (Figure 6)

Rectum DVHs show a small decrease at high doses ($\Delta V_{70Gy}$: -0.45%, p<0.001 and $\Delta V_{60Gy}$: -1.03%, p=0.007) and a more important reduction in the low dose region ($\Delta V_{30Gy}$: -3.17%, p=0.014).The mean dose to the rectum is significantly reduced (-128.30cGy, p=0.001). (Figure 7)

There are greater differences in bladder than in rectum, this is probably due to the fact that current clinical protocols impose more restrictive constraints on the latter.

## 4. Discussion

Complex delivery techniques imply more treatment planning time, more linac QA and also more patient specific QA time. Identifying potentially automatable tasks is therefore of great importance. If we focus on treatment planning, most of the times, during the treatment planning process, the same steps are followed, setting up a trial-and-error sequence that is susceptible of being automated. We developed a way to interact with Monaco TPS that allows the creation of automated workflows. The same approach can be followed to automatize other TPS since it does not need a specific API. Moreover, being able to programmatically evaluate the results of the optimization, allows a human-like interaction emulating an expert treatment planner.

We created site specific generic templates as a step towards homogenizing treatment planning. This work shows that this approach gives good results and guarantees minimum quality levels for prostate treatments. It generates plans with lower doses to organs at risk whilst negligibly reducing target coverage.

The web based calculation queue is now being used at Hospital General Universitario Gregorio Marañón, together with the Autoflow sequence, it is the current choice for simple prostate treatments. It has noticeably reduced treatment planning bottlenecks and has encouraged research.

There is ongoing work for extending its validity to other disease sites. We are aware that much effort has been devoted to automatize treatment planning over different centers, most of them with a great amount of resources available; this work shows that automatizing can be done in small steps and with little resources.

## 5. Conclusions

We developed a powerful toolkit to automate Monaco TPS without a specific API. An automatic workflow has been created that allows re-optimizations of radiotherapy plans. It has proved its validity for prostate treatments.

This work shows several ways to facilitate clinical routine: overnight calculations can boost planning timings, potentially reducing bottlenecks; if used together with an automatic planning sequence, it has the ability to homogenize the treatment planning process, giving more room for research and sometimes even improving the quality of the treatment plans.